\documentstyle[12pt]{article}

\def\hybrid{\topmargin -20pt    \oddsidemargin 0pt
        \headheight 0pt \headsep 0pt
        \textwidth 6.25in       
        \textheight 9.5in       
        \marginparwidth .875in
        \parskip 5pt plus 1pt   \jot = 1.5ex}
\def\cQ{{\cal Q}}
\def\cG{{\cal G}}
\def\cL{{\cal L}}
\def\cH{{\cal H}}
\def\ket#1{|{#1}\rangle}
\def\noi{\noindent}
\def\half{{1\over2}}
\def\baselinestretch{1.2}

\catcode`\@=11

\def\marginnote#1{}
\def\draftlabel#1{{\@bsphack\if@filesw {\let\thepage\relax
   \xdef\@gtempa{\write\@auxout{\string
      \newlabel{#1}{{\@currentlabel}{\thepage}}}}}\@gtempa
   \if@nobreak \ifvmode\nobreak\fi\fi\fi\@esphack}
        \gdef\@eqnlabel{#1}}
\def\@eqnlabel{}
\def\@vacuum{}
\def\draftmarginnote#1{\marginpar{\raggedright\scriptsize\tt#1}}

\def\draft{\oddsidemargin -.2truein
        \def\@oddfoot{\sl preliminary draft \hfil
        \rm\thepage\hfil\sl\today\quad\militarytime}
        \let\@evenfoot\@oddfoot \overfullrule 3pt
        \let\label=\draftlabel
        \let\marginnote=\draftmarginnote
   \def\@eqnnum{(\theequation)\rlap{\kern\marginparsep\tt\@eqnlabel}%
\global\let\@eqnlabel\@vacuum}  }

\def\preprint{\twocolumn\sloppy\flushbottom\parindent 2em
        \leftmargini 2em\leftmarginv .5em\leftmarginvi .5em
        \oddsidemargin -.5in    \evensidemargin -.5in
        \columnsep .4in \footheight 0pt
        \textwidth 10.in        \topmargin  -.4in
        \headheight 12pt \topskip .4in
88      \textheight 6.9in \footskip 0pt
        \def\@oddhead{\thepage\hfil\addtocounter{page}{1}\thepage}
        \let\@evenhead\@oddhead \def\@oddfoot{} \def\@evenfoot{} }

\def\numberbysection{\@addtoreset{equation}{section}
        \def\theequation{\thesection.\arabic{equation}}}

\def\underline#1{\relax\ifmmode\@@underline#1\else
        $\@@underline{\hbox{#1}}$\relax\fi}

\def\titlepage{\@restonecolfalse\if@twocolumn\@restonecoltrue
\onecolumn
     \else \newpage \fi \thispagestyle{empty}\c@page\z@
        \def\thefootnote{\fnsymbol{footnote}} }

\def\endtitlepage{\if@restonecol\twocolumn \else \newpage \fi
        \def\thefootnote{\arabic{footnote}}
        \setcounter{footnote}{0}}  

\catcode`@=12
\relax

\def\figcap{\section*{Figure Captions\markboth
        {FIGURECAPTIONS}{FIGURECAPTIONS}}\list
        {Figure \arabic{enumi}:\hfill}{\settowidth\labelwidth{Figure
999:}
        \leftmargin\labelwidth
        \advance\leftmargin\labelsep\usecounter{enumi}}}
 \relax
\def\tablecap{\section*{Table Captions\markboth
        {TABLECAPTIONS}{TABLECAPTIONS}}\list
        {Table \arabic{enumi}:\hfill}{\settowidth\labelwidth{Table
999:}
        \leftmargin\labelwidth
        \advance\leftmargin\labelsep\usecounter{enumi}}}
 \relax
\def\reflist{\section*{References\markboth
        {REFLIST}{REFLIST}}\list
        {[\arabic{enumi}]\hfill}{\settowidth\labelwidth{[999]}
        \leftmargin\labelwidth
        \advance\leftmargin\labelsep\usecounter{enumi}}}
 \relax
\makeatletter
\newcounter{pubctr}
\def\publist{\@ifnextchar[{\@publist}{\@@publist}}
\def\@publist[#1]{\list
        {[\arabic{pubctr}]\hfill}{\settowidth\labelwidth{[999]}
        \leftmargin\labelwidth
        \advance\leftmargin\labelsep
        \@nmbrlisttrue\def\@listctr{pubctr}
        \setcounter{pubctr}{#1}\addtocounter{pubctr}{-1}}}
\def\@@publist{\list
        {[\arabic{pubctr}]\hfill}{\settowidth\labelwidth{[999]}
        \leftmargin\labelwidth
        \advance\leftmargin\labelsep
        \@nmbrlisttrue\def\@listctr{pubctr}}}
 \relax
\makeatother
\newskip\humongous \humongous=0pt plus 1000pt minus 1000pt

\newif\ifdtup

\font\Scbig=cmss10 scaled\magstep1
\font\Scscr=cmss8 scaled\magstep1
\font\Scscrscr=cmss8
\newfam\Scfam
\textfont\Scfam=\Scbig
\scriptfont\Scfam=\Scscr
\scriptscriptfont\Scfam=\Scscrscr
\def\Sc{\fam\Scfam}

\relax
\hybrid
\def\lvm{\leavevmode\hbox to\parindent{\hfill}}

\def\thefootnote{\fnsymbol{footnote}}
\def\BE{\begin{equation}}
\def\EE{\end{equation}}
\def\BA{\begin{eqnarray}}
\def\EA{\end{eqnarray}}

\def\th{\theta}

\def\D{\Delta}

\def\tt{\bar\tau}
\def\Gz{\cG_0}
\def\Gn{$\Gz$}
\def\Qz{\cQ_0}
\def\Qn{$\Qz$}

\def\lvm{\leavevmode\hbox to\parindent{\hfill}}
\def\bar{\overline}
\def\req#1{(\ref{#1})}

\def\L{\left}
\def\R{\right}

\def\BE{\begin{equation}}
\def\EE{\end{equation} \vskip 0.30\baselineskip}
\def\BA{\begin{array}}
\def\EA{\end{array}}

\def\noi{\noindent}

\def\frac#1#2{{\textstyle{{#1}\over{#2}}}}
\def\half{{1\over2}}

\def\Kr#1{\delta_{{#1},0}}
\def\ket#1{|{#1}\rangle}

\def\cA{{\cal A}}
\def\cG{{\cal G}}
\def\cH{{\cal H}}

\def\cL{{\cal L}}

\def\cQ{{\cal Q}}

\def\cU{{\cal U}}

\def\open#1{\mbox{{\bf{#1}}}}

\def\oZ{{\open Z}}
\def\o1{{\open 1}}

\def\ctop{{\Sc c}}

\def\Ups{\Upsilon}
\def\kun{\ket{\Upsilon_{NS}}}
\def\kur{\ket{\Upsilon_{R}}}

\def\kc{\ket\chi}
\def\kcn{\ket{\chi_{NS}}}
\def\kcr{\ket{\chi_{R}}}

\def\svs{singular vectors}
\def\ie{{i.e.}}

\newif\ifold \oldtrue \def\new{\oldfalse}
\let\ssection=\section
\def\section{\setcounter{equation}{0}\ssection}

\begin{document}
\renewcommand{\theequation}{\thesection.\arabic{equation}}
\newcommand{\beq}{\begin{equation}}
\newcommand{\eeq}[1]{\label{#1}\end{equation}}
\newcommand{\ber}{\begin{eqnarray}}
\newcommand{\eer}[1]{\label{#1}\end{eqnarray}}
\begin{titlepage}
\begin{center}

\hfill IMAFF-FM-97/04, NIKHEF-97-048\\
\hfill HUTP-97/A055\\
\vskip .5in

{\large \bf Transmutations between Singular and Subsingular Vectors of the
N=2 Superconformal Algebras}
\vskip .8in

{\bf Matthias D\"orrzapf}$^{\,a}$ {\bf and Beatriz 
Gato-Rivera}$^{\,b,c\ }$\footnote{e-mail addresses: 
matthias@feynman.harvard.edu, {\ }bgato@imaff.cfmac.csic.es}\\

\vskip .3in

${\ }^a${\em Lyman Laboratory, Harvard University,
Cambridge, 02138 MA, USA}\\

\vskip .2in

${\ }^b${\em Instituto de Matem\'aticas y F\'\i sica Fundamental, 
CSIC,\\ Serrano 123,
Madrid 28006, Spain}\\ 

\vskip .2in

${\ }^c${\em NIKHEF, Kruislaan 409, NL-1098 SJ Amsterdam, The Netherlands}\\
 
\end{center}

\vskip .6in

\begin{center} {\bf ABSTRACT } \end{center} \begin{quotation} We present
subsingular vectors of the N=2 superconformal algebras other than the ones
which become singular in chiral Verma modules, reported recently by Gato-Rivera
and Rosado. We show that two large classes of singular vectors  of the
Topological algebra become subsingular vectors of the Antiperiodic NS  algebra
under the topological untwistings. These classes consist of BRST- invariant
singular vectors with relative charges $q=-2,-1$ and zero conformal  weight,
and no-label singular vectors with $q=0,-1$. In turn the resulting NS
subsingular vectors are transformed by the spectral flows into subsingular  and
singular vectors of the Periodic R algebra. We write down these singular  and
subsingular vectors starting from the topological singular vectors at  levels 1
and 2.

\end{quotation}
\vskip .6cm
March 1999

\end{titlepage}
\vfill
\eject
\def\baselinestretch{1.2}
\baselineskip 17 pt
\section{Introduction}\lvm

The N=2 Superconformal algebras provide the symmetries underlying the
N=2 strings \cite{Adem} \cite{Marcus}. These seem to be related 
to M-theory since many of the basic objects of M-theory 
are realized in the heterotic (2,1) N=2 strings \cite{Martinec}.
In addition, the topological version of the algebra is realized in the 
world-sheet of the bosonic string \cite{BeSe}, as well as in the world-sheet
of the superstrings \cite{BLNW}.  

Recently, subsingular vectors of the N=2 Superconformal algebras were
discovered in refs. \cite{BJI7}\cite{BJI6}. 
These are states which become singular
vectors (\ie\ highest weight null vectors) only in the quotient of a
Verma module by a submodule generated by singular vectors. Using a
different language subsingular vectors can be viewed as becoming
singular in incomplete Verma modules with constraints. 

The only explicit examples of N=2 subsingular vectors reported so far 
\cite{BJI7}\cite{BJI6} become singular in chiral Verma modules. That is,
they are singular either in chiral Verma modules of the Topological
algebra, or in chiral and antichiral Verma modules of the Antiperiodic
NS algebra, or in Verma modules of the Periodic R algebra built on the
Ramond ground states. The chiral Verma modules are incomplete, constrained
Verma modules which correspond to the quotient of complete Verma modules
by submodules generated by singular vectors (level-zero singular vectors
in the case of the Topological and R algebras, and level-1/2 singular
vectors in the case of the NS algebra).

In this paper we present N=2 subsingular vectors other than the ones which
become singular in chiral Verma modules. In section 2 we review briefly
some basic concepts and results related to the N=2 Superconformal algebras
in order to facilitate the reading of the main text. In section 3 we show
that two large classes of singular vectors of the
Topological algebra become subsingular vectors of the NS algebra under
the topological untwistings. These in turn are mapped to subsingular as 
well as to singular vectors of the R algebra by the spectral flows. In 
section 4 we write down these singular and subsingular vectors starting 
from the topological singular vectors at levels 1 
and 2. Section 5 is devoted to conclusions and final remarks.

\section{Basic Concepts}\lvm

\subsection{N=2 Superconformal algebras}\lvm                                 

The (non-topological) N=2 Superconformal algebras  
\cite{Adem}\cite{DiVPZ}\cite{SS}\cite{Kir1}\cite{LVW} can be expressed as

\BE\new\BA{lclclcl}
\L[L_m,L_n\R]&=&(m-n)L_{m+n}+{\ctop\over12}(m^3-m)\Kr{m+n}
\,,&\qquad&[H_m,H_n]&=
&{\ctop\over3}m\Kr{m+n}\,,\\
\L[L_m,G_r^\pm
\R]&=&\L({m\over2}-r\R)G_{m+r}^\pm
\,,&\qquad&[H_m,G_r^\pm]&=&\pm G_{m+r}^\pm\,,
\\
\L[L_m,H_n\R]&=&{}-nH_{m+n}\\
\L\{G_r^-,G_s^+\R\}&=&\multicolumn{5}{l}{2L_{r+s}-(r-s)H_{r+s}+
{\ctop\over3}(r^2-\frac{1}{4})
\Kr{r+s}\,,}\EA\label{N2alg}
\EE

\noi
where $L_m$ and $H_m$ are the spin-2 and spin-1 bosonic generators
corresponding to the stress-energy momentum tensor and the U(1) current,
respectively, and
$G_r^+$ and $G_r^-$ are the spin-3/2 fermionic generators. These 
are half-integer moded for the case of the Antiperiodic NS algebra, and
integer moded for the case of the Periodic R algebra. The eigenvalues of
the bosonic zero modes $(L_0, \,H_0)$ are the conformal weight and the
U(1) charge of the states. These are split conveniently as 
$(\D+l, \,h+q)$ for secondary states, where $l$ and $q$ are the level
and the relative charge of the state and $(\D, \,h)$ are the conformal
weight and U(1) charge of the primary on which the secondary is built.

Observe that we unify the notation for the $U(1)$
charge of the states of the NS algebra and the states of the R 
algebra since the $U(1)$ charges of the R states will be
denoted by $h$, instead of $h\pm \half$ which was commonly used in
the past \cite{bfk}, and their relative charges $q$ 
are defined to be integer, like for the NS states. 

There is also the twisted N=2 algebra for which the generators of the 
U(1) current are half-integer moded. As a consequence there is no
U(1) charge for this algebra.

\vskip .17in

The Topological N=2 Superconformal algebra reads \cite{DVV} 

\BE\new\BA{lclclcl}
\L[\cL_m,\cL_n\R]&=&(m-n)\cL_{m+n}\,,&\qquad&[\cH_m,\cH_n]&=
&{\ctop\over3}m\Kr{m+n}\,,\\
\L[\cL_m,\cG_n\R]&=&(m-n)\cG_{m+n}\,,&\qquad&[\cH_m,\cG_n]&=&\cG_{m+n}\,,
\\
\L[\cL_m,\cQ_n\R]&=&-n\cQ_{m+n}\,,&\qquad&[\cH_m,\cQ_n]&=&-\cQ_{m+n}\,,\\
\L[\cL_m,\cH_n\R]&=&\multicolumn{5}{l}{-n\cH_{m+n}+{\ctop\over6}(m^2+m)
\Kr{m+n}\,,}\\
\L\{\cG_m,\cQ_n\R\}&=&\multicolumn{5}{l}{2\cL_{m+n}-2n\cH_{m+n}+
{\ctop\over3}(m^2+m)\Kr{m+n}\,,}\EA\qquad m,~n\in\oZ\,.\label{topalgebra}
\EE

\noi
where the fermionic generators $\cQ_m$ and $\cG_m$ correspond
to the spin-1 BRST current and the spin-2 fermionic current, respectively,
\Qn\ being the BRST-charge. The eigenvalues of $(\cL_0, \, \cH_0)$ are
split, as before, as $(\D+l, \, h+q)$. This algebra is topological because 
the Virasoro generators can be expressed as $\cL_m=\half \{\cG_m, \Qz \}$.
This implies, as is well known, that the correlators of the fields do
not depend on the metric.

\subsection{Topological twists}\lvm

The Topological algebra \req{topalgebra} can be viewed as a rewriting of
the algebra \req{N2alg} using one of the two topological twists:

\BE\new\BA{rclcrcl}
\cL^{(1)}_m&=&\multicolumn{5}{l}{L_m+\half(m+1)H_m\,,}\\
\cH^{(1)}_m&=&H_m\,,&{}&{}&{}&{}\\
\cG^{(1)}_m&=&G_{m+\half}^+\,,&\qquad &\cQ_m^{(1)}&=&G^-_{m-\half}
\,,\label{twa}\EA\EE

\noi
and

\BE\new\BA{rclcrcl}
\cL^{(2)}_m&=&\multicolumn{5}{l}{L_m-\half(m+1)H_m\,,}\\
\cH^{(2)}_m&=&-H_m\,,&{}&{}&{}&{}\\
\cG^{(2)}_m&=&G_{m+\half}^-\,,&\qquad &
\cQ_m^{(2)}&=&G^+_{m-\half}\,,\label{twb}\EA\EE                

\noi
which we denote as $T_{W1}$ and $T_{W2}$, respectively. Observe that the 
two twists are mirrored under the exchange $H_m \to -H_m$, $G^+_r  
\leftrightarrow G^-_r$. In particular $(G^{+}_{1/2}, G^{-}_{-1/2})$
results in $(\cG^{(1)}_0, \cQ^{(1)}_0)$, while
$(G^{-}_{1/2}, G^{+}_{-1/2})$ gives $(\cG^{(2)}_0, \cQ^{(2)}_0)$, so
that the zero mode 
\Gn\ corresponds to the positive modes $G^{\pm}_{1/2}$ of the NS algebra.
As a result, all the highest weight
(h.w.) states of the NS algebra (primary states and
singular vectors) are transformed, under $T_{W1}$ and $T_{W2}$, into
states of the Topological algebra annihilated by \Gn\, which are also
h.w. states, as the reader can easily verify. The other way around, all
the h.w. states of the Topological algebra 
annihilated by \Gn\ are transformed under $T_{W1}$ and 
$T_{W2}$ into h.w. states of the NS algebra. 
The zero mode \Qn , in turn, corresponds to the negative modes 
$G^{\mp}_{-1/2}$ of the NS algebra. Therefore the topological states
annihilated by \Qn\ become antichiral and chiral states of the NS algebra
(annihilated by $G^-_{-1/2}$ and by $G^+_{-1/2}$, respectively)
under the twists $T_{W1}$ and $T_{W2}$.

\subsection{Spectral flows}\lvm

The spectral flows $\,\cU_{\th}$ and $\cA_{\th}$ are 
one-parameter families of transformations providing a continuum of
isomorphic N=2 Superconformal algebras. The `usual' spectral flow 
$\cU_{\th}$ \cite{SS}\cite{LVW}\cite{BJI4}\cite{Be1} is even, given by 

\BE\new\BA{rclcrcl}
\cU_\th \, L_m \, \cU_\th^{-1}&=& L_m
 +\th H_m + {\ctop\over 6} \th^2 \delta_{m,0}\,,\\
\cU_\th H_m \, \cU_\th^{-1}&=&H_m + {\ctop\over3} \th \delta_{m,0}\,,\\
\cU_\th \, G^+_r \, \cU_\th^{-1}&=&G_{r+\th}^+\,,\\
\cU_\th \, G^-_r \, \cU_\th^{-1}&=&G_{r-\th}^-\,,\
\label{spfl} \EA\EE

\noi 
satisfying $\, \cU^{-1}_\th = \cU_{(-\th)}$. For $\th=0$ it is just the
identity operator, \ie\ $\, \cU_0={\bf 1}$. It transforms the $(L_0, H_0)$
eigenvalues, \ie\ the conformal weight and the
U(1) charge, $(\Delta, h)$ of a given state as
 $(\Delta-\th h +{\ctop\over6} \th^2, \,h- {\ctop\over3} \th)$. From this 
one gets straightforwardly that the level $l$ of any secondary state 
changes to $l-\th q$, while the relative charge $q$ remains equal.

The spectral flow $\cA_{\th}$ \cite{BJI4}\cite{Be1} is odd, given by

\BE\new\BA{rclcrcl}
\cA_\th \, L_m \, \cA_\th^{-1}&=& L_m
 +\th H_m + {\ctop\over 6} \th^2 \delta_{m,0}\,,\\
\cA_\th H_m \, \cA_\th^{-1}&=&- H_m - {\ctop\over3} \th \delta_{m,0}\,,\\
\cA_\th \, G^+_r \, \cA_\th^{-1}&=&G_{r-\th}^-\,,\\
\cA_\th \, G^-_r \, \cA_\th^{-1}&=&G_{r+\th}^+\,,\
\label{ospfl} \EA\EE

\noi
satisfying $\cA_{\th}^{-1} = \cA_{\th}$. It is therefore an involution. 
The odd spectral flow $\cA_{\th}$ is `quasi' mirror symmetric to 
the even spectral flow $\cU_{\th}$,  
under the exchange $H_m \to -H_m$, $ G_r^+ \leftrightarrow G_r^-$ and 
$\th \to -\th$, and it is in fact the only fundamental spectral flow 
since it generates the latter \cite{Be1}.
For $\th=0$ it is the mirror map, \ie\ $\cA_0={\cal M}$.
It transforms the $(L_0, H_0)$ eigenvalues of the states as
$(\Delta+\th h +{\ctop\over6} \th^2, - h - {\ctop\over3} \th)$.
The level $l$ of the secondary states changes 
to $l + \th q$, while the relative charge $q$ reverses its sign.

For half-integer values of $\th$ the two spectral flows interpolate 
between the NS algebra and the R algebra. In particular, for 
$\th = \pm 1/2$ the h.w. states of the NS algebra 
are transformed into h.w. states of the R
algebra with helicities $(\mp)$(\ie\ annihilated by $G^-_0$ and $G^+_0$,
respectively). As a result the NS singular vectors are transformed into R
singular vectors with helicities $(\mp)$ built on R primaries with the
same helicities. 

By performing the topological twists $T_{W1}$ \req{twa} and 
$T_{W2}$ \req{twb} on the spectral flows one obtains the topological 
spectral flows \cite{BJI3}\cite{Be1}.

\section{Transmutation between N=2 Singular and Subsingular Vectors}\lvm

In what follows the singular vectors of the Topological algebra, the 
NS algebra and the R algebra will be denoted as $\kc$, $\kcn$ and $\kcr$,
respectively. The states which are not singular, like the subsingular
vectors, will be denoted as $\ket{\Ups}$, $\kun$ and $\kur$, respectively.

\subsection{Topological singular vectors}\lvm

It has recently been shown \cite{BJI6}\cite{DB2} that the singular
vectors of the Topological algebra
can be classified in 29 different types in complete Verma
modules, and in 4 different types in chiral Verma modules, regarding
the relative charge $q$ and the BRST-invariance properties of the
singular vectors and of the primaries on which they are built. Namely,
there are ten types of 
topological singular vectors built on \Gn-closed primaries
$\ket{\D, \, h}^G$ (\ie\ annihilated by \Gn ),
ten types of topological singular vectors built on \Qn-closed primaries
$\ket{\D, \, h}^Q$ (\ie\ annihilated by \Qn ), 
four types built on chiral primaries 
$\ket{0, \, h}^{G,Q}$ (annihilated by both \Gn\ and \Qn ),
and nine types built on no-label primaries $\ket{0, \, h}$ (which cannot be 
expressed as linear combinations of \Gn-closed, \Qn-closed and chiral 
primaries). The no-label and the chiral topological states (primaries as 
well as secondaries) have zero conformal weight, as deduced
from the anticommutator $\{ \Gz, \Qz \} = 2 \cL_0$.

The \Qn-closed and no-label h.w. states (primaries or singular vectors) 
are transformed, under $T_{W1}$ \req{twa} and $T_{W2}$ \req{twb}, into 
states of the NS algebra which are not h.w. states since they are not
annihilated by one of the modes $G^{\pm}_{1/2}$ (because the topological
state is not annihilated by \Gn ). The \Gn-closed and chiral h.w. 
states, however, are transformed into h.w. states of the NS algebra.
In particular the \Gn-closed primaries 
$\ket{\D, \, h}^G$ give rise to the topological Verma modules 
which are transformed into the NS Verma modules, whereas
the chiral primaries $\ket{0, \, h}^{G,Q}$ generate topological 
chiral (incomplete) Verma modules which are transformed into antichiral 
and chiral (incomplete) Verma modules of the NS algebra under $T_{W1}$ 
and $T_{W2}$, respectively.

In what follows we will restrict our attention to the topological
singular vectors built on \Gn-closed primaries $\ket{\D, \, h}^G$.
The ten different types of singular vectors in this case are shown in 
the table\footnote{The results of this table were conjectured in ref.
\cite{BJI6} and have been rigorously proved in ref. \cite{DB2}.}
below \cite{BJI6}\cite{DB2}.  The chiral and no-label singular 
vectors have zero conformal weight, therefore they satisfy $\D+l=0$. 

\vskip .35in
\BE \begin{tabular}{r|l l l l}
{\ } & $q=-2$ & $q=-1$ & $q=0$ & $q=1$ \\
\hline \\
\Gn-closed & $-$ & $\kc^{(-1)G}_l$ & $\kc^{(0)G}_l$ & $\kc^{(1)G}_l$ \\
\Qn-closed & $\kc^{(-2)Q}_l$ & $\kc^{(-1)Q}_l$ & $\kc^{(0)Q}_l$ & $-$ \\
chiral & $-$ & $\kc^{(-1)G,Q}_l$ & $\kc^{(0)G,Q}_l$ & $-$\\
no-label & $-$ & $\kc^{(-1)}_l$ & $\kc^{(0)}_l$ & $-$\\
 \end{tabular} \label{table}  \EE

\vskip .25in

An important observation is that chiral singular vectors can be viewed
as particular cases of \Gn-closed singular vectors and/or as particular
cases of \Qn-closed singular vectors, which `become' chiral 
when the conformal weight turns out to be zero. In particular,
the singular vectors of types $\kc^{(0)Q}_l$ and $\kc^{(-1)G}_l$
in table \req{table} always become chiral when the 
conformal weight is zero \cite{Be2}\cite{DB2}. 

The spectrum of $(\D,\, h)$ corresponding to the \Gn-closed primaries 
which contain singular vectors in their Verma modules
has been derived in ref. \cite{BJI6} for the \Gn-closed, for the 
\Qn-closed, and (partially) for the chiral singular vectors. For
the latter it has been derived completely in ref. \cite{DB4} 
together with the spectrum corresponding to no-label singular vectors.

\subsection{Transmutation between topological and NS singular and 
subsingular vectors}\lvm

In this subsection we will identify two large classes of topological \svs ,
built on \Gn-closed primaries,
which become subsingular vectors of the NS
algebra under the topological untwistings. These classes consist of
\Qn-closed (BRST-invariant) \svs\ with zero conformal weight, which only
exist for relative charges $q=-2,-1$, and no-label \svs , which only 
exist for $q=0,-1$.

\vskip .2in

\noi
\Qn-{\bf closed singular vectors}

Let us consider a \Qn-closed topological singular vector $\kc^{(q)Q}_l$, 
from table \req{table}, with zero
conformal weight. The action of \Gn\ on this singular
vector produces another singular vector with zero conformal weight,
at level $l$, with relative charge $q+1$, and annihilated by \Gn\ . 
Moreover, since $\D+l=0$ this singular vector is also annihilated by
\Qn\ :

\BE \Qz\,\Gz\, \kc^{(q)Q}_l = - \Gz\,\Qz\, \kc^{(q)Q}_l + 
2\, \cL_0 \, \kc^{(q)Q}_l = 2(\D+l)\, \kc^{(q)Q}_l = 0 \, .  \EE

As a result, $\kc^{(q+1)G,Q}_l = \Gz \, \kc^{(q)Q}_l$ is a chiral 
singular vector which cannot `come back' to the singular vector 
$\,\kc^{(q)Q}_l$ by acting with the algebra (the chiral singular vector
is therefore a level-zero secondary singular vector with respect to
this one). An important observation now is that chiral singular vectors
of type $\,\kc^{(1)G,Q}_l$ do not exist, according to table \req{table}.
As a consequence the singular vectors of type $\,\kc^{(0)Q}_l$
with zero conformal weight, built on \Gn-closed primaries, are absent too,   
`becoming' all of them chiral; that is, of type $\kc^{(0)G,Q}_l$ instead
(as $\Gz \, \kc^{(0)Q}_l = 0$).

Let us repeat this procedure from the viewpoint of the NS algebra.
That is, let us untwist every step using $T_{W1}$ \req{twa} 
and $T_{W2}$ \req{twb}. To start, the \Qn-closed topological 
singular vector $\,\kc^{(q)Q}_l$, built on the primary $\ket{-l,\,h}^G$, 
is not transformed into
singular vectors of the NS algebra but into mirrored antichiral and chiral
states $\,\kun^{(\pm q)\mp}_{l-q/2}$, built on the mirrored NS primaries
$\ket{-l-h/2, \pm h}$,
which are not annihilated by $\,G^{\pm}_{1/2}$ (the upper signs 
using $T_{W1}$ and the lower signs using $T_{W2}$, the superscript $(-)$ 
for antichiral and $(+)$ for chiral). These states 
are annihilated, however, by all other positive modes of the NS algebra,
so that $\,G^{\pm}_{1/2} \, \kun^{(\pm q)\mp}_{l-q/2}\,$ produces
truly singular vectors of the NS algebra which are antichiral and chiral
respectively. That is,  
$\,\kcn^{(q+1)-}_{l-(q+1)/2} =  G^{+}_{1/2} \, \kun^{(q)-}_{l-q/2}\,$ is an 
antichiral NS singular vector (annihilated by $G^-_{-1/2}$), in the Verma
module $\,V_{NS}(\ket{-l-h/2, \, h})$, whereas 
$\,\kcn^{(-q-1)+}_{l-(q+1)/2} = G^{-}_{1/2} \, \kun^{(-q)+}_{l-q/2}\,$ 
is the mirrored chiral NS singular vector (annihilated by $G^+_{-1/2}$), in
the mirrored Verma module $V_{NS}(\ket{-l-h/2, \, -h})$. From 
these singular vectors it
is not possible to reach back the states $\,\kun^{(\pm q)\mp}_{l-q/2}$,
respectively, which are located therefore outside the (incomplete) 
Verma modules built on the singular vectors.

Hence the antichiral and chiral NS states $\,\kun^{(\pm q)\mp}_{l-q/2}$, 
obtained by untwisting the \Qn-closed topological \svs\ with zero 
conformal weight, are 
mirrored subsingular vectors of the NS algebra in the mirrored Verma 
modules $\,V_{NS}(\ket{-l-h/2, \, \pm h})$. 
Once the quotients of the Verma modules by the corresponding 
singular vectors are performed, by setting these to zero:
$\,G^{\pm}_{1/2} \, \kun^{(\pm q)\mp}_{l-q/2} = 0$, the subsingular
vectors become singular because they recover the only missing
h.w. condition.

\vskip .17in The spectrum of U(1) charges $h$ corresponding to the \Gn-closed
primaries $\ket{-l, \, h}^G$ which contain \Qn-closed singular vectors at level
$l$ in their Verma modules can be deduced from the general formulae given in
refs. \cite{BJI6}\cite{DB4}. As to the chiral singular vectors, the
corresponding spectrum has been given in ref. \cite{DB4}. One obtains the
following results.  The topological Verma module $V_T(\ket{-l, \, h}^G)$
contains two singular vectors at   level $l$ of the types $\, \kc^{(-2)Q}_l\ $
and $\, \kc^{(-1)G,Q}_l\ $ for  \BE  h = {t \over 2} (1+l) + 1 \ , 
\label{chh1} \EE  \noi and two singular vectors at level $l$ of the types  $\,
\kc^{(-1)Q}_l\ $ and $\, \kc^{(0)G,Q}_l\ $ for  \BE  h = {t (1-r) - s \over 2}
\ , \label{chh0} \EE  \noi where $t= {3-\ctop \over3}$ and $(r,\, s)$ are two
positive  integers ($s$ even) such that $l={r s \over2}$. For the discrete
values \BE t = - {s \over n} \,, \ \ \ \  n = 1,...,r \,, \label{chht} \EE \noi
however, $\, \kc^{(-1)Q}_l\ $ also becomes chiral, i.e. of type  $\,
\kc^{(-1)G,Q}_l\ $ instead.

\vskip .3in
\noi
{\bf No-label singular vectors}

Now let us consider a no-label singular vector $\kc^{(q)}_l$ from
table \req{table}. This vector cannot be expressed as a linear combination
of \Gn-closed, \Qn-closed and chiral singular vectors, and
has zero conformal weight necessarily,
\ie\ $\D+l=0$. The action of \Gn\ on $\kc^{(q)}_l$ produces a
\Gn-closed singular vector $\kc^{(q+1)G}_l$ with zero conformal weight
at the same level. The action of \Qn\ on $\kc^{(q+1)G}_l$ now produces
a chiral singular vector $\kc^{(q)G,Q}_l = \cQ_0 \cG_0 \kc^{(q)}_l$, 
for the same reasons as in the previous case. 

{\ }From the viewpoint of the NS algebra $\,\kc^{(q)}_l$ does not correspond
to a singular vector since it is not annihilated by \Gn\ and therefore
is not annihilated by one of the modes $\,G^{\pm}_{1/2}$ under the 
untwistings, as happened with the \Qn-closed topological singular vectors
in the previous case. Let us denote as $\,\kun^{(\pm q)}_{l-q/2}$ the 
mirrored states, built on the mirrored primaries $\ket{-l-h/2, \pm h}$,
obtained under the untwistings of $\kc^{(q)}_l$. 
The action of 
$\,G^{\pm}_{1/2}$ on these states produce NS singular vectors
$\,\kcn^{(\pm (q+1))}_{l-(q+1)/2}= G^{\pm}_{1/2} \kun^{(\pm q)}_{l-q/2}$. 
{\ }From these, however, it is not possible
to reach the previous non-singular states $\,\kun^{(\pm q)}_{l-q/2}$
because the action of $G^{\mp}_{-1/2}$ on the singular vectors produces
antichiral and chiral NS singular vectors instead (the untwistings 
of $\kc^{(q)G,Q}_l$). That is, $G^{\mp}_{-1/2} 
\kcn^{(\pm (q+1))}_{l-(q+1)/2}= \kcn^{(\pm q)\mp}_{l-q/2}$.

Hence the untwistings of the no-label topological singular vectors 
$\,\kc^{(q)}_l$ produce mirrored subsingular vectors 
$\,\kun^{(\pm q)}_{l-q/2}$
of the NS algebra which become singular when the  
singular vectors $\,\kcn^{(\pm (q+1))}_{l-(q+1)/2} = 
G^{\pm}_{1/2} \, \kun^{(\pm q)}_{l-q/2}$ are set to zero.
Observe that the antichiral and chiral NS 
singular vectors $\,\kcn^{(\pm q)\mp}_{l-q/2}$ also ``go away" because 
they are descendant, secondary singular vectors of the singular
vectors $\,\kcn^{(\pm (q+1))}_{l-(q+1)/2}$.

There is also the possibility of acting with \Qn\ on $\kc^{(q)}_l$. In this
case one obtains a \Qn-closed topological singular vector 
$\,\kc^{(q-1)Q}_l$ with zero
conformal weight, like the ones we analyzed before.
Therefore $\,\kc^{(q)}_l$ and $\,\kc^{(q-1)Q}_l = \Qz \, \kc^{(q)}_l$, 
both in the topological Verma module $V_T(\ket{-l,\, h}^G)$, are
transformed into subsingular vectors of the NS algebra under
the topological untwistings. Namely, $\kc^{(q)}_l$ is transformed into
non-chiral mirrored NS subsingular vectors of the types 
$\,\kun^{(\pm q)}_{l-q/2}$, in the mirrored Verma modules  
$V_{NS}(\ket{-l-h/2,\,\pm h})$,
while $\,\kc^{(q-1)Q}_l$ is transformed into antichiral and chiral mirrored
NS subsingular vectors of the types $\,\kun^{(\pm (q-1))\mp}_{l-(q-1)/2} 
= G^{\mp}_{-1/2} \, \kun^{(\pm q)}_{l-q/2}$, in the same mirrored Verma
modules.

\vskip .17in
The no-label singular vectors are very scarce since they only exist 
for $t={2 \over r}$, i.e. $\ \ctop={3r-6 \over r}\ $ \cite{DB4}. Thus 
at level 1 they only exist for $\ctop=-3$ and at level 2 they 
only exist for $\ctop=-3$ and $\ctop=0$. The spectrum of U(1) charges $h$ 
corresponding to the \Gn-closed primaries $\ket{-l, \, h}^G$ 
which contain no-label singular vectors at level $l = {rs \over 2}$
in their Verma modules is \cite{DB4} 
\BE  h = {1 \over r} - {s \over 2} -1 \,, \ \ \ t={2 \over r} \,. 
\label{nlh} \EE

\subsection{Transmutation between NS and R singular and subsingular
vectors}\lvm

The spectral flows $\, \cU_{1/2}$ and $\cA_{1/2}$, \req{spfl} 
and \req{ospfl}, transform the h.w. states of the NS algebra 
(primaries and singular vectors) into h.w. states of the R algebra 
annihilated by $G^-_0$, denoted as helicity $(-)$ states. 
The spectral flows $\, \cU_{-1/2}$ and $\cA_{-1/2}$, 
on the contrary, transform the h.w. states of the NS algebra into h.w. 
states of the R algebra annihilated by $G^+_0$, denoted as helicity $(+)$ 
states. 

Now we will investigate the spectral flow transformations, into R states,
of the NS subsingular vectors analyzed in the last subsection.  
For this purpose let us inspect the transformations of the ``missing" 
h.w. conditions $\,G^{\pm}_{1/2} \, \kun \neq 0$ and the transformations 
of the (anti)chirality conditions $\,G^{\mp}_{-1/2} \, \kun = 0$,
which correspond to the NS subsingular vectors obtained under the 
untwisting of the \Qn-closed topological singular vectors with zero
conformal weight (whereas the untwisting of the no-label topological 
singular vectors gives non-chiral NS subsingular vectors).
The transformations of $G^{\pm}_{1/2}$ and $G^{\mp}_{-1/2}$ 
under $\cU_{\pm 1/2}$ and $\cA_{\pm 1/2}$ are:

\BE \BA{rclrrcl} 
\cU_{1/2} \, G^+_{1/2} \, \cU_{-1/2} &=& G^+_1 \ \ & \qquad &
\cU_{1/2} \, G^-_{-1/2} \, \cU_{-1/2} &=& G^-_{-1}\\
\cU_{1/2} \, G^-_{1/2} \, \cU_{-1/2} &=& G^-_0 \ \  & \qquad &
\cU_{1/2} \, G^+_{-1/2} \, \cU_{-1/2} &=& G^+_{0} \EA \EE

\BE \BA{rclcrcl} 
\cA_{1/2} \, G^+_{1/2} \, \cA_{1/2} &=& G^-_0 \ \ & \qquad &
\cA_{1/2} \, G^-_{-1/2} \, \cA_{1/2} &=& G^+_{0}\\
\cA_{1/2} \, G^-_{1/2} \, \cA_{1/2} &=& G^+_1 \ \ & \qquad &
\cA_{1/2} \, G^+_{-1/2} \, \cA_{1/2} &=& G^-_{-1} \EA \EE

\BE \BA{rclcrcl} 
\cU_{-1/2} \, G^+_{1/2} \, \cU_{1/2} &=& G^+_0 \ \ & \qquad &
\cU_{-1/2} \, G^-_{-1/2} \, \cU_{1/2} &=& G^-_{0}\\
\cU_{-1/2} \, G^-_{1/2} \, \cU_{1/2} &=& G^-_1 \ \ & \qquad &
\cU_{-1/2} \, G^+_{-1/2} \, \cU_{1/2} &=& G^+_{-1} \EA \EE

\BE \BA{rclcrcl} 
\cA_{-1/2} \, G^+_{1/2} \, \cA_{-1/2} &=& G^-_1 \ \ & \qquad &
\cA_{-1/2} \, G^-_{-1/2} \, \cA_{-1/2} &=& G^+_{-1}\\
\cA_{-1/2} \, G^-_{1/2} \, \cA_{-1/2} &=& G^+_0 \ \ & \qquad &
\cA_{-1/2} \, G^+_{-1/2} \, \cA_{-1/2} &=& G^-_{0} \EA \EE

{\ }From the expressions on the left one ``reads" the missing conditions   
(h.w. or helicity) of the corresponding R states, whereas
the expressions on the right give the 
constraints associated to each case (when they apply). 

We see that
under $\cU_{1/2}$ the missing h.w. condition $G^+_{1/2} \, \kun \neq 0$
on the NS subsingular vectors results in the missing h.w. condition
$G^+_{1} \, \kur^- \neq 0$ on the corresponding helicity $(-)$ R states. 
The missing h.w.
condition $G^-_{1/2} \, \kun \neq 0$, however, results in the ``missing
helicity" $(-)$ on the R states. The chirality constraint 
$G^+_{-1/2} \, \kun =0$, when it applies, compensates the loss of
the helicity $(-)$ providing the helicity $(+)$ to the corresponding 
states. The antichirality constraint $G^-_{-1/2} \, \kun =0$ is just
transformed into the constraint $G^-_{-1} \, \kur^- =0$ on the R states.

Under $\cA_{1/2}$ the missing h.w. condition $G^+_{1/2} \, \kun \neq 0$
leads to the ``missing helicity" $(-)$, although the antichirality
constraint compensates (when it applies) providing a helicity $(+)$
to the corresponding R state. The missing h.w. condition
$G^-_{1/2} \, \kun \neq 0$, on the other hand, results in the missing
h.w. condition $G^+_{1} \, \kur^- \neq 0$ and the chirality constraint
gives the constraint $G^-_{-1} \, \kur^- =0$ on the R states.

Similar remarks apply to the transformations of the missing h.w.
conditions and the transformations 
of the (anti)chirality constraints under $\cU_{-1/2}$
and $\cA_{-1/2}$. In these cases the resulting missing h.w. condition 
on the helicity $(+)$ R states is $G^-_{1} \, \kur^+ \neq 0$, the resulting 
``missing helicity" is $(+)$, and the resulting constraint on the R states 
(when it applies) is $G^+_{-1} \, \kur^+ =0$.

{\ }From this analysis one deduces straightforwardly the following. 
First, the two mirrored antichiral and chiral NS subsingular vectors
obtained by untwisting the \Qn-closed topological singular vectors are
mapped, under all four spectral flows $\cU_{\pm 1/2}$ and $\cA_{\pm 1/2}$, 
to one R singular vector, with $(+)$ or $(-)$ helicity, and one
state which is not singular and satisfies a constraint. Second,
the two mirrored non-chiral NS subsingular vectors
obtained by untwisting the no-label topological singular vectors are
mapped, under all four spectral flows $\cU_{\pm 1/2}$ and $\cA_{\pm 1/2}$, 
to one R singular vector without any helicity (annihilated only by 
the positive modes of the R algebra), and one state which is not singular. 

Now let us show that the R state which is not singular, in each case,
is actually a subsingular vector of the R algebra. Let us take the spectral
flow $\,\cU_{1/2}$, eq. \req{spfl}. The NS subsingular vectors with the
missing h.w. condition $\,G^+_{1/2}\, \kun \neq 0$ are of the types
$\,\kun^{(q)-}_{l-q/2}$ and $\,\kun^{(q)}_{l-q/2}$, depending on whether
they come from the untwisting of the \Qn-closed topological singular
vectors or from the untwisting of the no-label topological singular vectors.
In both cases they are located in the Verma modules
$\,V_{NS}(\ket{-l-h/2, \, h})$ (with different spectrum of $h$ for each
type). These NS subsingular vectors are transformed
by $\,\cU_{1/2}$ into R states
of the types $\,\kur^{(q)-}_{l-q}$ (some of them with the constraint
$G^-_{-1} \, \kur = 0$), in the Verma modules 
$\,V_R(\ket{-l-h+\ctop/24, \, h-\ctop/6}^-)$, which
satisfy all but one of the h.w. conditions: 
$\ G^+_1 \, \kur^{(q)-}_{l-q} \neq 0$. But $\ G^+_1 \, \kur^{(q)-}_{l-q}$
is a singular vector of type $\,\kcr^{(q+1)-}_{l-q-1}$, as one can easily
verify, which cannot reach back the state $\,\kur^{(q)-}_{l-q}$.

To see this let us analyze the action of $G^-_{-1}$ on the singular vector
$\,\kcr^{(q+1)-}_{l-q-1} = G^+_1 \, \kur^{(q)-}_{l-q}$:
\BE G^-_{-1} G^+_1 \, \kur^{(q)-}_{l-q} = 
- G^+_{1} G^-_{-1} \, \kur^{(q)-}_{l-q} + 
2 (L_0 + H_0 + \ctop/8) \, \kur^{(q)-}_{l-q} = 
- G^+_{1} G^-_{-1} \, \kur^{(q)-}_{l-q} \EE
This result vanishes for the states with the constraint
$\,G^-_{-1} \, \kur^{(q)-}_{l-q} = 0$, which originate from the \Qn-closed
topological singular vectors. For the states without this constraint,
which originate from the no-label topological singular vectors, 
$\,G^-_{-1} G^+_1 \, \kur^{(q)-}_{l-q}$ is a singular vector of type
$\,\kcr^{(q)-}_{l-q}$ annihilated by $G^-_{-1}$, as one can easily verify.

Hence the spectral flow $\cU_{1/2}$ transforms the NS subsingular vectors 
of the types $\,\kun^{(q)-}_{l-q/2}$ and $\,\kun^{(q)}_{l-q/2}$,
in the Verma modules $V_{NS}(\ket{-l-h/2, \, h})$, into R subsingular
vectors of the types $\,\kur^{(q)-}_{l-q}$,
in the Verma modules $V_{R}(\ket{-l-h+\ctop/24, \, h-\ctop/6}^-)$.

Similar reasonings and results apply to the spectral flow mappings
$\cU_{-1/2}$ and $\cA_{\pm 1/2}$ on the NS subsingular vectors;  
that is, the R state which is not singular, in each case, is a 
subsingular vector of the R algebra.

\newcommand{\bea}{\begin{eqnarray}}
\newcommand{\eea}{\end{eqnarray}}
\def\noi {\noindent}
\newcommand{\nn}{\nonumber}

\section{Results from the Topological Singular Vectors 
at Levels 1 and 2}\lvm

In this section we will present the explicit expressions for the singular 
and subsingular vectors that we have analyzed in the previous section, 
starting from the topological singular vectors at levels 1 and 2. We will
follow the usual parametrization $\ctop=3-3t$. 

\subsection{Level 1}\lvm

\noi
\Qn-{\bf closed topological singular vectors versus NS subsingular vectors}

According to table \req{table}, the $\cQ_0$-closed topological 
singular vectors $\kc_{l}^{(q)Q}$
built on $\cG_0$-closed primaries $\ket{\D,\,h}^G$
only exist for relative charges $q=$ $0$, $-1$, $-2$. 
At zero conformal weight $\D +l=0$, however,
there are no singular vectors of the type $\kc_{l}^{(0)Q}$ because all
of them ``become" chiral, \ie\ of type $\kc^{(0)G,Q}_l$, as we explained
in section 3. The negatively charged singular vectors exist for both 
allowed values of $q$. As one can deduce from expressions \req{chh1}
and \req{chh0}, at level $1$ these vectors can be found for $h=-1$ 
in the case $q=-1$ and for $h=1+t$ in the case $q=-2$:
 
\BE \BA{rcl}
\kc_1^{(-1)Q} &=& \Big\{ t \cL_{-1} \cQ_0 -2 
               \cQ_{-1} -2 \cH_{-1} \cQ_{0} \Big\}
               \ket{-1,-1}^G \\
\kc_1^{(-2)Q} &=& \cQ_{-1} \cQ_0 \ket{-1,1+t}^G 
\EA \EE

Under the action of $\cG_0$, these vectors are mapped to 
chiral singular vectors $\kc_1^{(0)G,Q}$ and $\kc_1^{(-1)G,Q}$
respectively:
\BE \BA{lccl}
\kc_1^{(0)G,Q}& = & \cG_0 \kc_1^{(-1)Q} & = 
(t+2) \Big\{ \cG_{-1} \cQ_0 -2 \cL_{-1} \Big\} \ket{-1,-1}^G \\
\kc_1^{(-1)G,Q}& = & \cG_0 \kc_1^{(-2)Q} & = 
 2 \Big\{ \cL_{-1} \cQ_0 + \cH_{-1} \cQ_0 + \cQ_{-1} \Big\}
 \ket{-1,1+t}^G 
\EA \EE
\noi
Observe that for $t=-2$ $\ \kc_1^{(-1)Q}\ $ is annihilated by \Gn , i.e.
it becomes chiral of type $\ \kc_1^{(-1)G,Q}\ $ instead, as predicted 
by eq. \req{chht} (level 1 corresponds to $r=1, \, s=2$).
Untwisting $\kc_{1}^{(-1)Q}$ and $\kc_{1}^{(-2)Q}$ using $T_{W1}$ \req{twa}
leads to the antichiral NS subsingular vectors:
\BE \BA{lcl}
\kun_{\frac{3}{2}}^{(-1)-} 
&=& \Big\{ -2G^-_{-\frac{3}{2}} +t L_{-1}G^-_{-\frac{1}{2}}
               -2H_{-1}G^-_{-\frac{1}{2}} \Big\} \ket{-\frac{1}{2},-1} \, ,\\
\kun_{2}^{(-2)-} &=& G^-_{-\frac{3}{2}} G^-_{-\frac{1}{2}}
\ket{-\frac{3+t}{2},1+t}
\, .
\EA \label{eq:ch2} \EE

  Using $T_{W2}$ \req{twb} instead one gets mirrored chiral NS
subsingular vectors \ie\ which differ from the antichiral ones by 
the exchange 
$H_m \to - H_m$, $G^\pm_r \to G^\mp_r$, $h\to -h$ and $q\to -q$.

The action of $G^+_{1/2}$ 
maps $\,\kun_{3/2}^{(-1)-}$ and $\,\kun_2^{(-2)-}$ to the antichiral
NS singular vectors $\,\kcn_1^{(0)-}$ (for $t\neq -2$) and
 $\,\kcn_{3/2}^{(-1)-}$, respectively:
\BE \BA{lccl}
\kcn_1^{(0)-}& = & G^{+}_{\frac{1}{2}} \kun_{\frac{3}{2}}^{(-1)-} &=
 (t+2)\Big\{ -2L_{-1} + G^+_{-\frac{1}{2}}
  G^-_{-\frac{1}{2}} \Big\} \ket{-\frac{1}{2},-1} \, ,\\
\kcn_{\frac{3}{2}}^{(-1)-}& 
=& G^{+}_{\frac{1}{2}} \kun_2^{(-2)-}=& 2\Big\{ G^-_{-\frac{3}{2}}
                                    +L_{-1}G^-_{-\frac{1}{2}} 
                                    +H_{-1}G^-_{-\frac{1}{2}} \Big\} 
                                    \ket{-\frac{3+t}{2},1+t} \, .
\EA \EE
The action of any other positive modes on the vectors
$\,\kun_{3/2}^{(-1)-}$ 
and $\,\kun_2^{(-2)-}$ vanishes identically. Thus these vectors
are subsingular with respect to the corresponding singular vectors which,
being antichiral, cannot reach the subsingular vectors back.
As $\,\kcn_1^{(0)-}= 0\,$ for $t=-2$
($\ctop=9$), the vector $\,\kun_{3/2}^{(-1)-}$ 
turns out to be singular in this particular case. 

It is worth mentioning that the NS algebra does not have singular
vectors with relative charges different from $|q| = 0, 1$ \cite{Doerr2}.
This fact is important for understanding the $N=2$
character formulae. Most surprisingly is therefore that there are
subsingular vectors with relative charges $|q|=2$.

\vskip .2in
\noi
{\bf No-label topological singular vectors versus NS subsingular vectors}

Let us now investigate the NS subsingular vectors obtained by untwisting the 
topological no-label singular vectors at level $1$. 
As table \req{table} shows, the no-label singular vectors on \Gn-closed
primaries only exist for relative charges $q=0 , -1$. As explained in
section 3, the restrictions 
on this kind of vectors are very strict, such that they exist 
just for some discrete values of $t$:  
$t={2 \over r}$, i.e. $\ \ctop={3r-6 \over r}\ $.

At level $1$ there are no-label singular vectors only for $t=2 \ (\ctop=-3)$  
as $r=1$. These vectors and the \Gn-closed singular vectors obtained 
by the action of $\cG_0$ are given by:

\BE \BA{rcl}
\kc_1^{(0)} &=& \Big\{ \cH_{-1} - \cL_{-1} \Big\} \ket{-1,-1,t=2}^G \, ,\\
\kc_1^{(1)G}\ = \cG_0 \kc_1^{(0)} &=& -2 \cG_{-1} \ket{-1,-1,t=2}^G \, ,\\
\kc_1^{(-1)} &=& \Big\{ \cL_{-1}\cQ_0 +2 \cH_{-1}\cQ_0 \Big\} 
                    \ket{-1,3,t=2}^G \, ,\\
\kc_1^{(0)G} = \cG_0 \kc_1^{(-1)} 
&=& -\Big\{ \cG_{-1} \cQ_0 +2\cL_{-1} +4\cH_{-1} \Big\} \ket{-1,3,t=2}^G \, .  
\EA  \label{nlg} \EE

By untwisting the no-label singular vectors using  $T_{W1}$ one finds
the NS subsingular vectors 
 
\BE \BA{lcl}
\kun_1^{(0)} &=& \Big\{ H_{-1}-L_{-1} \Big\} \ket{-\frac{1}{2},-1,t=2} \\
\kun_{\frac{3}{2}}^{(-1)} &=& 
\Big\{ L_{-1}G^-_{-\frac{1}{2}}+2H_{-1}G^-_{-\frac{1}{2}}
                   \Big\} \ket{-\frac{5}{2},3,t=2} \, . 
\EA \EE

This kind of subsingular vector is always accompanied by an antichiral
 singular vector at the same level and with the same charge, which
corresponds to the untwisting of the chiral topological singular vector. 
Thus, the NS subsingular vectors obtained by untwisting the no-label
topological singular vectors also have different representations, like
the latter. The vectors 
$\,\kun_1^{(0)}$ and $\,\kun_{3/2}^{(-1)}$ are subsingular with respect
to the singular vectors
\BE \BA{rcl}
\kcn_{\frac{1}{2}}^{(1)} =& G^{+}_{\frac{1}{2}} \kun_1^{(0)} 
&= -2G^+_{-\frac{1}{2}} \ket{-\frac{1}{2},-1,t=2} \, ,\\
\kcn_1^{(0)} =& G^{+}_{\frac{1}{2}} \kun_{\frac{3}{2}}^{(-1)} 
&= -\Big\{ 2L_{-1}+4H_{-1}
+G^+_{-\frac{1}{2}}G^-_{-\frac{1}{2}} \Big\}
\ket{-\frac{5}{2},3,t=2} \, , \EA \EE
\noi 
respectively, obtained by untwisting the \Gn-closed singular vectors in
 \req{nlg}.

By untwisting the no-label singular vectors in \req{nlg} using $T_{W2}$
one obtains the mirror-symmetric NS subsingular vectors which are 
subsingular with respect to the mirror-symmetric NS singular vectors.

\vskip .2in
\noi
{\bf NS subsingular vectors versus R singular and subsingular vectors}

We now use the spectral flows $\,\cU_{\pm 1/2}$ and $\cA_{\pm 1/2}$ in order
to transform the NS subsingular vectors $\kun_{3/2}^{(-1)-}$ ($t\neq -2$),
$\kun_{2}^{(-2)-}$, $\kun_1^{(0)}$, and $\kun_{3/2}^{(-1)}$
into R singular and subsingular vectors. Using $\,\cU_{1/2}$ and $\cA_{-1/2}$
we obtain subsingular vectors in the R algebra as well. Using $\,\cU_{-1/2}$
and $\cA_{1/2}$, however, we obtain R singular vectors instead.
The R subsingular vectors obtained using $\,\cU_{1/2}$ are 

\BE \BA{lccl}
\kur_{2}^{(-1)-}& =& \cU_{1/2}\kun_{\frac{3}{2}}^{(-1)-} &= 
                \Big\{ \! -2G^-_{-2} +tL_{-1}G^-_{-1} 
                +(\frac{t}{2}-2)H_{-1}G^-_{-1}
                \Big\} \ket{\frac{1-t}{8},\frac{t-3}{2}}^- \nn \\
\kur_{3}^{(-2)-}& =& \cU_{1/2}\kun_{2}^{(-2)-} &= 
                G^-_{-2}G^-_{-1} \ket{-\frac{15+9t}{8},\frac{1+3t}{2}}^- 
                \nn \\
\kur_1^{(0)-}& =& \cU_{1/2}\kun_1^{(0)} &= 
                \Big\{ \frac{1}{2}H_{-1} -L_{-1} \Big\} 
                \ket{-\frac{1}{8},-\frac{1}{2},t=2}^-   \\
\kur_{2}^{(-1)-}& =& \cU_{1/2}\kun_{\frac{3}{2}}^{(-1)} &= 
                \Big\{ L_{-1}G^-_{-1} +\frac{5}{2}H_{-1}G^-_{-1} \Big\}
                \ket{-\frac{33}{8},\frac{7}{2},t=2}^-  \ , \label{h-sv}
\EA \EE
\noi
where the helicity $(-)$ of the R vectors (subsingular and primaries) 
indicate that they are annihilated by $G_0^-$. $\,\kur_{2}^{(-1)-}$ and 
 $\,\kur_{3}^{(-2)-}$ are in addition also annihilated by $G_{-1}^-$.

Using the spectral flow $\cU_{-1/2}$ we obtain 
the helicity $(-)$ and the no-helicity R singular vectors:
\BE \BA{lccl}
\kcr_{1}^{(-1)-}& =& \cU_{-1/2}\kun_{\frac{3}{2}}^{(-1)-} &= 
                \Big\{ \! - \! 2G^-_{-1} +tL_{-1}G^-_{0} 
                -(\frac{t}{2}+2)H_{-1}G^-_{0}
                \Big\} \ket{\frac{-7-t}{8},\frac{-1-t}{2}}^+ \nn \\
\kcr_{1}^{(-2)-}& =& \cU_{-1/2}\kun_{2}^{(-2)-} &= 
                G^-_{-1}G^-_{0} \ket{-\frac{7+t}{8},\frac{3+t}{2}}^+ 
                \nn \\
\kcr_1^{(0)}& =& \cU_{-1/2}\kun_1^{(0)} &= 
                \Big\{ \frac{3}{2}H_{-1} -L_{-1} \Big\} 
                \ket{-\frac{9}{8},-\frac{3}{2},t=2}^+ \\
\kcr_{1}^{(-1)}& =& \cU_{-1/2}\kun_{\frac{3}{2}}^{(-1)} &= 
                \Big\{ L_{-1}G^-_{0} +\frac{3}{2}H_{-1}G^-_{0} \Big\}
                \ket{-\frac{9}{8},\frac{5}{2},t=2}^+  \ , \label{h-nh}
\EA  \EE
\noi
where the helicity $(+)$ of the R primaries indicate that they are annihilated
by $G^+_0$. This result is due to the fact that $\cU_{-1/2}$ transforms 
$G^+_{1/2}$ into $G^+_0$, so that instead of a missing h.w. condition one
has the missing helicity $(+)$ for the resulting R singular vectors.
The antichirality of the NS subsingular vectors $\,\kun_{3/2}^{(-1)-}$
and $\,\kun_{2}^{(-2)-}$ compensates this loss as it is converted into
helicity $(-)$ for the resulting R singular vectors. 

The corresponding results for $\cA_{\pm 1/2}$ are simply mirror-symmetric
to the above expressions. To be precise, $\cA_{1/2}$ produces helicity $(+)$
and no-helicity R singular vectors mirrored to the ones in eq. \req{h-nh},
whereas $\cA_{-1/2}$ produces helicity $(+)$ R subsingular vectors mirrored
to the ones in eq. \req{h-sv}.  

Two important remarks are now in order. First, the R singular
vectors $\kcr_{l}^{(-2)-}$ with relative charge $q=-2$, 
built on the helicity $(+)$ primaries $\ket{\D, h}^+$, can be regarded
equivalently, provided $\D \neq {\ctop / 24}$, as singular vectors 
with $q=-1$ built on the helicity $(-)$ primaries
$\ket{\D, h-1}^- = G_0^- \ket{\D, h}^+ $.
Second, `no-helicity' R singular vectors have never been considered in the 
literature before\footnote{Between the first version and the present
version of this paper we have considered `no-helicity' R singular vectors
also in refs. \cite{DB2} and \cite{DB4}.}. 
These R singular vectors, which are analogous
to the no-label topological singular vectors, are permitted by 
the R algebra provided the conformal weight satisfies $\D + l = \ctop /24$,
as the anticommutator $\{G^+_0 , G^-_0\} = 2 L_0 - \ctop / 12$ shows.
Like the no-label topological singular vectors, the no-helicity R 
singular vectors only exist for $t={2 \over r}$. 
The existence of no-helicity R singular vectors does not contradict the 
determinant formula because they are always accompanied by 
helicity $(+)$ and helicity $(-)$ singular vectors at the same level, 
obtained by the action of $G^+_0$ and $G^-_0$. These two issues are
explained in detail in ref. \cite{DB4}.

\subsection{Level 2}\lvm

\noi
\Qn-{\bf closed topological singular vectors versus NS subsingular vectors}

At level $2$ the $\cQ_0$-closed topological singular vectors with zero 
conformal weight, built on $\cG_0$-closed primaries, are the following.
As deduced from eqs. \req{chh1} and \req{chh0},
there are singular vectors with $q=-1$ for $h=-2$ and $h=-1-\frac{t}{2}$,
which we label by $a$ and $b$ respectively, and there are singular
vectors with $q=-2\,$ for $\,h=1+\frac{3}{2}t$. They are given by:

\bea
\kc_2^{(-1)Q,a} &=& \Big\{ 8t \cL_{-1} \cQ_{-1} + 6t \cL_{-1} \cH_{-1} \cQ_0
                 -4 (2+t) \cH_{-2} \cQ_0 
                 -8 \cH_{-1}^2 \cQ_0 -4(t+4) \cQ_{-2} \nn \\
&&                 + t \cG_{-1}\cQ_{-1}\cQ_0 -16 \cH_{-1}\cQ_{-1} 
                 -t^2 \cL_{-1}^2 \cQ_0 \Big\} \ket{-2,-2}^G \nn \\
\kc_2^{(-1)Q,b} &=& \Big\{ -(3t+2) \cL_{-1}\cQ_{-1} 
                   -t(2+\frac{t}{2}) \cL_{-1}\cH_{-1} \cQ_0
                   +(1+\frac{t}{2})(2+t) \cH_{-2}\cQ_0  \\ 
&&                 +(1-\frac{t^2}{4})(2+t) \cL_{-2}\cQ_0
                   +\frac{t^2}{2} \cL_{-1}^2 \cQ_0 
                   -\frac{4+2t-t^2}{4} \cG_{-1}\cQ_{-1}\cQ_0 \nn \\
&&                 +(2+t) \cH_{-1}^2 \cQ_0
                   +2(2+t) \cH_{-1}\cQ_{-1} 
                   +(2+\frac{t}{2})(2+t) \cQ_{-2} \Big\} 
                   \ket{-2,-1-\frac{t}{2}}^G \nn \\
\kc_2^{(-2)Q} \ &=& \Big\{ 2 \cL_{-1}\cQ_{-1}\cQ_0 +4\cH_{-1}\cQ_{-1}\cQ_0
                   +(2-t) \cQ_{-2}\cQ_0 \Big\} \ket{-2,1+\frac{3}{2}t}^G \nn 
\eea

The singular vector $\,\kc_2^{(-1)Q,a}\,$ becomes chiral for $t=-4$
(it corresponds to $r=1,\ s=4$) and $\,\kc_2^{(-1)Q,b}\,$
becomes chiral for $t=-1,-2$ (it corresponds to $r=2,\ s=2$),
according to the result \req{chht}.
The corresponding NS subsingular vectors, obtained by untwisting using
$T_{W1}$, together with the NS singular vectors
with respect to which they are subsingular, are the following:
 
\bea
\kun_{\frac{5}{2}}^{(-1)-,a}\  &=& 
\Big\{ -4(t+4)G^-_{-\frac{5}{2}} +8tL_{-1}G^-_{-\frac{3}{2}}
                 -4(2+t)H_{-2}G^-_{-\frac{1}{2}}  \nn \\ &&
                 +6tL_{-1}H_{-1}G^-_{-\frac{1}{2}} 
                 -16H_{-1}G^-_{-\frac{3}{2}}
                 -8H_{-1}^2G^-_{-\frac{1}{2}} 
                 -t^2L_{-1}^2G^-_{-\frac{1}{2}} \nn \\ &&
                 +t G^+_{-\frac{1}{2}}
                 G^-_{-\frac{3}{2}} G^-_{-\frac{1}{2}} \Big\} \ket{-1,-2} 
                  \, , \;\; t\neq -4\, ,  \nn  \\
\kcn_2^{(0)-,a} \ \ \ &=& 
G^{+}_{\frac{1}{2}} \kun_{\frac{5}{2}}^{(-1)-,a} = 
                         (t+4) \Big\{ -8L_{-2} +4H_{-2}
                         -8L_{-1}H_{-1} +4tL_{-1}^2 \nn \\
&&                         +4G^+_{-\frac{1}{2}}G^-_{-\frac{3}{2}}
                         +4H_{-1}G^+_{-\frac{1}{2}}G^-_{-\frac{1}{2}}
                         -2tL_{-1}G^+_{-\frac{1}{2}}G^-_{-\frac{1}{2}}
                         \Big\} \ket{-1,-2} \, , \nn
\eea
\bea
\kun_{\frac{5}{2}}^{(-1)-,b} &=& 
                 \Big\{ (2+t) (2+\frac{t}{2}) G^-_{-\frac{5}{2}}
                 -(3t+2)L_{-1}G^-_{-\frac{3}{2}}
                 +\frac{2+t}{2} (1+t+\frac{t^2}{4}) H_{-2}G^-_{-\frac{1}{2}}
                  \nn \\
&&         +(2+t)(1-\frac{t^2}{4})L_{-2}G^-_{-\frac{1}{2}}
                 +2(2+t) H_{-1}G^-_{-\frac{3}{2}}
                 -t(2-\frac{t}{2})L_{-1}H_{-1}G^-_{-\frac{1}{2}}
                  \nn \\
&& 
                 +(2+t)H_{-1}^2G^-_{-\frac{1}{2}}
                 +\frac{t^2}{2}L_{-1}^2G^-_{-\frac{1}{2}}
                 -(1+\frac{t}{2}-\frac{t^2}{2})G^+_{-\frac{1}{2}}
                 G^-_{-\frac{3}{2}}G^-_{-\frac{1}{2}} \Big\} \nn \\ &&
                 \ket{-\frac{3}{2}+\frac{t}{4},-1-\frac{t}{2}} 
                  \, , \;\; t\neq -1,-2\, , \\
\kcn_2^{(0)-,b} \ &=& 
G^{+}_{\frac{1}{2}} \kun_{\frac{5}{2}}^{(-1)-,b} = (1+t)(2+t) 
                     \Big\{ tL_{-2} 
                     -\frac{t}{2}H_{-2} -2L_{-1}^2 \nn \\ &&
                     +(1-\frac{t}{2})G^+_{-\frac{3}{2}}
                     G^-_{-\frac{1}{2}}  
                     -G^+_{-\frac{1}{2}}G^-_{-\frac{3}{2}}  
                     +2L_{-1}H_{-1} 
                     +L_{-1}G^+_{-\frac{1}{2}}G^-_{-\frac{1}{2}}
                     \nn \\ &&
                     -H_{-1}G^+_{-\frac{1}{2}}G^-_{-\frac{1}{2}}
                     \Big\} \ket{-\frac{3}{2}+\frac{t}{4},-1-\frac{t}{2}} 
                     \, , \nn
\eea
\bea
\kun_3^{(-2)-} &=& \!\! \Big\{ (2-t)G^-_{-\frac{5}{2}}G^-_{-\frac{1}{2}}
               +2L_{-1}G^-_{-\frac{3}{2}}G^-_{-\frac{1}{2}}
               +4H_{-1}G^-_{-\frac{3}{2}}G^-_{-\frac{1}{2}} 
   \Big\} \ket{-\frac{5}{2}-\frac{3}{4}t,1+\frac{3}{2}t} , \nn \\
\kcn_{\frac{5}{2}}^{(-1)-} &=& \!\!
G^{+}_{\frac{1}{2}} \kun_3^{(-2)-} = \Big\{ 4(2-t)G^-_{-\frac{5}{2}}
                     +2(2-t)L_{-2}G^-_{-\frac{1}{2}}
                     -(3t+2)H_{-2}G^-_{-\frac{1}{2}} \nn \\ &&
                     +8L_{-1}G^-_{-\frac{3}{2}}  
                     +16H_{-1}G^-_{-\frac{3}{2}}
                     -2G^+_{-\frac{1}{2}}G^-_{-\frac{3}{2}}G^-_{-\frac{1}{2}}
                     +4L_{-1}^2G^-_{-\frac{1}{2}}
                     +8H_{-1}^2G^-_{-\frac{1}{2}} \nn \\ &&
                     +12L_{-1}H_{-1}G^-_{-\frac{1}{2}} 
                     \Big\} 
                     \ket{-\frac{5}{2}-\frac{3}{4}t,1+\frac{3}{2}t} \, . \nn
\eea

Observe that for $t=-4$ the subsingular vector 
$\kun_{5/2}^{(-1)-,a}\ $ becomes singular 
(since $\kcn_2^{(0)-,a}=0$)  and similarly for
$t=-1,-2$ the subsingular vector $\kun_{5/2}^{(-1)-,b}\ $
becomes singular (since $\kcn_2^{(0)-,b}=0$).
By untwisting using $T_{W2}$ one finds the mirror-symmetric NS subsingular
vectors which are subsingular with respect
to the mirror-symmetric NS singular vectors. 

\vskip .2in
\noi
{\bf No-label topological singular vectors versus NS subsingular vectors}

At level $2$ the no-label topological singular vectors exist only for $t=1$ 
and $t=2$ ($\ctop = 0$ and $\ctop=-3$). They are the following:  

\bea
\kc_2^{(0)}\ &=& \Big\{ 4\cH_{-1}^2 -6\cL_{-1}\cH_{-1} -2\cL_{-2}
                    +\cL_{-1}\cG_{-1}\cQ_0 -\cG_{-1}\cQ_{-1} +8\cH_{-2}\nn \\
&&            +\cG_{-2}\cQ_0 \Big\} \ket{-2,-2,t=2}^G  \nn \\ 
\kc_2^{(0)} \ &=& \Big\{ 24 \cH_{-1}^2 -10\cH_{-1}\cG_{-1}\cQ_0 -48
 \cL_{-1}^2
                    +26\cL_{-1}\cG_{-1}\cQ_0  \nn \\
&&            -20\cG_{-1}\cQ_{-1} +36\cH_{-2}
                    +19 \cG_{-2}\cQ_0 \Big\} \ket{-2,-\frac{3}{2},t=1}^G \\
\kc_2^{(-1)} &=& \Big\{ 12 \cL_{-1}\cH_{-1}\cQ_0 -4\cL_{-1}\cQ_{-1}
                     -23 \cH_{-2}\cQ_0 -6\cL_{-2}\cQ_0 
                     +17\cH_{-1}^2\cQ_0  \nn \\
&&             +\cL_{-1}^2\cQ_0 +\cG_{-1}\cQ_{-1}\cQ_0 -2\cH_{-1}\cQ_{-1}
                     -6\cQ_{-2} \Big\} \ket{-2,4,t=2}^G \nn \\
\kc_2^{(-1)} &=& \Big\{ \cL_{-1}\cQ_{-1} +9\cL_{-1}\cH_{-1}\cQ_0 
                     -\frac{21}{2}\cH_{-2}\cQ_0
                     -\frac{3}{2}\cL_{-2}\cQ_0 +11 \cH_{-1}^2\cQ_0 
                     +\cL_{-1}^2\cQ_0 
                      \nn \\
&&             -\cG_{-1}\cQ_{-1}\cQ_0 +10\cH_{-1}\cQ_{-1} -
                  \frac{5}{2}\cQ_{-2}
                     \Big\} \ket{-2,\frac{5}{2},t=1}^G  \nn
\eea

The untwisting of these topological singular vectors, using $T_{W1}$, 
leads to the following NS subsingular vectors, given together with the
NS singular vectors with respect to which they are subsingular:

\bea
\kun_2^{(0)} \  &=& \Big\{ -2 L_{-2} +9H_{-2} +4H_{-1}^2
                    -6L_{-1}H_{-1} -G^+_{-\frac{1}{2}}G^-_{-\frac{3}{2}}
                    +G^+_{-\frac{3}{2}}G^-_{-\frac{1}{2}} \nn \\ &&
                    +L_{-1}G^+_{-\frac{1}{2}}G^-_{-\frac{1}{2}}
                    \Big\} \ket{-1,-2,t=2} \, , \nn \\
\kcn_{\frac{3}{2}}^{(1)} \  &=& 
G^{+}_{\frac{1}{2}} \kun_2^{(0)} = 12 \Big\{ L_{-1}G^+_{-\frac{1}{2}}
                        -H_{-1}G^+_{-\frac{1}{2}} \Big\} \ket{-1,-2,t=2} \, ,
                           \nn \\
\kun_2^{(0)} \ &=& \Big\{ 36 H_{-2} +24 H_{-1}^2 -48 L_{-1}^2
                    -20G^+_{-\frac{1}{2}}G^-_{-\frac{3}{2}}
                    +19G^+_{-\frac{3}{2}}G^-_{-\frac{1}{2}}  \nn \\
&&            -10H_{-1}G^+_{-\frac{1}{2}}G^-_{-\frac{1}{2}}
                    +26L_{-1}G^+_{-\frac{1}{2}}G^-_{-\frac{1}{2}}
                    \Big\} \ket{-\frac{5}{4},-\frac{3}{2},t=1} \, ,\\
\kcn_{\frac{3}{2}}^{(1)} \ &=& 
G^{+}_{\frac{1}{2}} \kun_2^{(0)} = 24 \Big\{ G^+_{-\frac{3}{2}}
                    +2L_{-1}G^+_{-\frac{1}{2}} -2H_{-1}G^+_{-\frac{1}{2}}
                    \Big\} \ket{-\frac{5}{4},-\frac{3}{2},t=1} \, ,\nn \\
\kun_{\frac{5}{2}}^{(-1)} &=& 
                     \Big\{ -6G^-_{-\frac{5}{2}} -4L_{-1}G^-_{-\frac{3}{2}}
                     -20H_{-2}G^-_{-\frac{1}{2}} -6L_{-2}G^-_{-\frac{1}{2}}
                     +12L_{-1}H_{-1}G^-_{-\frac{1}{2}} \nn \\ && 
                     -2H_{-1}G^-_{-\frac{3}{2}}
                     +17H_{-1}^2G^-_{-\frac{1}{2}} +L_{-1}^2G^-_{-\frac{1}{2}}
                     +G^+_{-\frac{1}{2}}G^-_{-\frac{3}{2}}G^-_{-\frac{1}{2}}
                     \Big\} \ket{-4,4,t=2} \, , \nn \\
\kcn_2^{(0)} \  &=& 
G^{+}_{\frac{1}{2}} \kun_{\frac{5}{2}}^{(-1)} = 6\Big\{ 2L_{-2} +11H_{-2} 
                     -G^+_{-\frac{1}{2}}G^-_{-\frac{3}{2}}
                     -12H_{-1}^2 \nn 
                     -10L_{-1}H_{-1} \nn \\ &&
                     -2L_{-1}^2 +3G^+_{-\frac{3}{2}}G^-_{-\frac{1}{2}}
                     -2L_{-1}G^+_{-\frac{1}{2}}G^-_{-\frac{1}{2}}
                     -4H_{-1}G^+_{-\frac{1}{2}}G^-_{-\frac{1}{2}}
                     \Big\} \ket{-4,4,t=2} \nn \, ,\\
\kun_{\frac{5}{2}}^{(-1)} &=& \Big\{ -\frac{5}{2}G^-_{-\frac{5}{2}}
                     +L_{-1}G^-_{-\frac{3}{2}}
                     -\frac{39}{4}H_{-2}G^-_{-\frac{1}{2}}
                     -\frac{3}{2}L_{-2}G^-_{-\frac{1}{2}} 
                     +9L_{-1}H_{-1}G^-_{-\frac{1}{2}} \nn \\ &&
                     +10H_{-1}G^-_{-\frac{3}{2}} 
                     +11H_{-1}^2G^-_{-\frac{1}{2}}
                     +L_{-1}^2G^-_{-\frac{1}{2}}
                     -G^+_{-\frac{1}{2}}G^-_{-\frac{3}{2}}G^-_{-\frac{1}{2}}
                     \Big\} \ket{-\frac{13}{4},\frac{5}{2},t=1} , \nn \\
\kcn_2^{(0)}\ &=& 
G^{+}_{\frac{1}{2}} \kun_{\frac{5}{2}}^{(-1)} = \Big\{ \frac{23}{2} H_{-2}
                     +L_{-2} +\frac{15}{2}G^+_{-\frac{3}{2}}G^-_{-\frac{1}{2}}
                     -24H_{-1}^2 -14L_{-1}H_{-1} \nn \\ &&
                     -2L_{-1}^2  
                     -5G^+_{-\frac{1}{2}}G^-_{-\frac{3}{2}}
                     -11H_{-1}G^+_{-\frac{1}{2}}G^-_{-\frac{1}{2}}
                     -5L_{-1}G^+_{-\frac{1}{2}}G^-_{-\frac{1}{2}}
                     \Big\} \ket{-\frac{13}{4},\frac{5}{2},t=1} \, .\nn
\eea

By untwisting using $T_{W2}$ one finds the mirror-symmetric NS subsingular
vectors which are subsingular with respect to the mirror-symmetric NS 
singular vectors. 

The spectral flow transformations of these NS subsingular vectors into
singular and subsingular vectors of the R algebra 
we leave to the interested reader. 

\section{Conclusions and Final Remarks}\lvm 

We have considered singular and subsingular vectors of the N=2
Superconformal algebras. The latter are non-highest weight null
vectors located outside the (incomplete) Verma submodules
built on certain singular vectors, becoming singular 
(i.e. highest weight) after taking the quotient of 
the Verma module by the submodule generated by the singular vector.

We have shown that two large classes of singular vectors of the
Topological algebra, built on \Gn-closed primaries, correspond to
 subsingular vectors of the NS algebra and to singular and subsingular
vectors of the R algebra. These classes consist of $Q_0$-closed 
(BRST-invariant) singular vectors with relative charges $q=-2,-1$ and
zero conformal weight, and no-label singular vectors with $q=0,-1$.
 This provides one more step towards the derivation of the
correct $N=2$ embedding diagrams and towards understanding the $N=2$
character formulae. 

We have also shown the existence of subsingular vectors 
of the NS algebra with $|q|=2$. This has significant
implications for the embedding diagrams as subsingular vectors
should be included in them. (It was believed so far that embedding
diagrams for the $N=2$ algebras contained only vectors of charge 
$|q|=0,1$).

In addition we have written down, for the first time, examples of 
R singular vectors with relative charge $|q|=2$ and examples of
R singular vectors without any helicity. These types of R singular vectors  
have been overlooked in the early literature and only recently they have 
been paid some attention \cite{DB2}\cite{DB4}. 

Besides the classes of subsingular vectors 
of the $N=2$ algebras presented in this work, 
there is, so far, only one other class of subsingular
vectors of the $N=2$ algebras known. These are subsingular 
vectors of the Topological, NS and R algebras, 
which become singular in chiral Verma modules 
\cite{BJI7}\cite{BJI6}. A chiral Verma module is the quotient of
a complete Verma module by a submodule generated by a lowest-level  
singular vector. One could think of constructing, from the beginning, 
 quotient modules of Verma modules with higher-level singular vectors.
These quotient spaces would possibly lead to subsingular vectors in exactly
the same way as chiral Verma modules do. As a matter of fact
we have already constructed further classes of subsingular
vectors in this way, what will be presented in a forthcoming publication. 

\vskip .3in 

\centerline{\bf Acknowledgements}

We would like to thank S. Hwang, A. Kent, J. Navarro-Salas and
A. Tiemblo for valuable comments.
M.D. is supported by a DAAD fellowship and in part by
NSF grant PHY-92-18167.    

\vskip .2in
\noi
{\bf Note}

\noi
{\small In the recent paper `The Structure of Verma Modules over the  
N=2 Superconformal Algebra', Comm. Math. Phys. 195 (1998) 129,
(hep-th/9704111) by A.M. Semikhatov and I.Yu. Tipunin, the authors
claim to present a complete classification of N=2 subsingular vectors.
However, the only explicit examples known at that time, due to
Gato-Rivera and Rosado \cite{BJI7}\cite{BJI6}, do not fit into that 
classification. In addition, the classification is based on several 
misleading assumptions, the most relevant being the following.
First, the authors claim that there are only two types of submodules,
overlooking no-label singular vectors which do not fit into their
submodules and create a different type of submodules themselves
(no-label singular vectors were discovered and examples given in
ref. \cite{BJI6}). Second, they claim to have constructed some null
states, called non-conventional singular vectors, which generate
maximal submodules such that there are not subsingular vectors 
outside of them. We have counterexamples that show that these
states do not generate maximal submodules as one can find subsingular
vectors outside of them. (In fact, the existence of no-label singular
vectors already proves that the states given by the authors do not 
generate maximal submodules).
Third, the authors confuse, all along the
paper, the determinant formula for the Topological N=2 algebra (which
was unknown at that time and has been written down just recently 
\cite{DB4}) with the determinant formula for the Neveu-Schwarz N=2
algebra. In addition they believe that the determinant formula identify
all possible submodules (what has never been proved for the N=2 algebras).
Finally, the authors claim that all the results found for the 
Verma modules of the Topological N=2 algebra hold for the Verma modules
of the Neveu-Schwarz and the Ramond N=2 algebras. The work presented 
by us in this paper shows precisely the contrary: two large classes of
singular vectors of the Topological algebra do not correspond to
singular vectors of the Neveu-Schwarz N=2 algebra, but to subsingular
vectors. (As was analysed in refs. \cite{BJI7}\cite{BJI6}, only the 
topological singular vectors annihilated by \Gn , built on primaries
annihilated also by \Gn , correspond to singular vectors of the 
Neveu-Schwarz N=2 algebra.
Some other topological singular vectors correspond simply to
null descendants of NS singular vectors).}

\end{document}